\documentclass[aps,a4paper,showpacs,preprint,onecolumn]{revtex4}%
\usepackage{amsfonts}
\usepackage{amsmath}
\usepackage{amssymb}
\usepackage{graphicx}%
\begin{document}
\title{Spin-dependent tunneling through a symmetric semiconductor barrier: the
Dresselhaus effect\\ }
\author{L. G. Wang, Wen Yang}
\author{Kai Chang}
\altaffiliation{Correspondence should be sent to: kchang@red.semi.ac.cn}

\affiliation{NLSM, Institute of Semiconductors, Chinese Academy of Sciences, P. O. Box 912,
Beijing 100083, China}
\author{K. S. Chan}
\affiliation{Department of Physics and Materials Science, City University of Hong Kong,
Kowloon, Hong Kong, China}


\begin{abstract}
Spin-dependent tunneling through a symmetric semiconductor barrier is studied
including the $k^{3}$ Dresselhaus effect. The spin-dependent transmission of
electron can be obtained analytically. By comparing with previous work(Phys.
Rev. B \textbf{67}. R201304 (2003) and Phys. Rev. Lett. \textbf{93}. 056601
(2004)), it is shown that the spin polarization and interface current are
changed significantly by including the off-diagonal elements in the current
operator, and can be enhanced considerably by the Dresselhaus effect in the
contact regions.

\end{abstract}
\maketitle

Recently electron spin in semiconductors has attracted a rapidly growing
interest due to its potential application in spintronics devices. To
successfully incorporate spin into existing semiconductor technology, one has
to overcome technical difficulties such as efficient spin-polarized injection,
transport, control and manipulation, as well as measurement of spin
polarization. The injection of spin-polarized electrons from ferromagnetic
metals into semiconductors has low efficiency, less than 1\%, because of a
large resistivity mismatch between ferromagnetic and semiconductor
materials\cite{GS}. Rashba proposed that this problem could be solved by
inserting a tunneling barrier at the metal-semiconductor
interface\cite{Rashba}. Asymmetric nonmagnetic semiconductor barriers are also
used in the construction of spin filters\cite{Lee}. This effect is caused by
the interface-induced Rashba spin-orbit coupling\cite{Yu} and can be quite
significant for resonant tunneling through asymmetric double-barrier
structure\cite{Liu}. Very recently, a multichannel field-effect spin-barrier
selector was investigated theoretically utilizing the Rashba and Dresselhaus
effects\cite{GEM}. A considerable spin polarization and an interesting
"tunneling spin-galvanic" effect were found in the tunneling of electron
through a single symmetric barrier utilizing the Dresselhaus effect in the
barrier\cite{INY,SAT}. However, the off-diagonal elements in the current
operator, the contribution from the Dresselhaus effect, are neglected in these
works. These off-diagonal element in the currents operator could lead to the
significant correction of the spin-dependent transmission, especially in a
thin barrier case.

In this paper, we investigate theoretically the spin-dependent tunneling
through a single symmetric barrier. The barrier and contacts consist of a
zinc-blende-structure semiconductor lacking the inversion symmetry. It is
shown that the spin-dependent tunneling and the electric current in the plane
of the interfaces are different from the previous studies\cite{INY}. This
difference is significant in thin barrier case and disappears gradually with
increasing the thickness of the barrier. It is interesting to notice that the
spin polarization and the interface current $j_{\parallel}$ are enhanced by
including the Dresselhaus effect in the contact regions.

We consider the transmission of an electron with initial wave vector
$\boldsymbol{k}=(\boldsymbol{k}_{\shortparallel},k_{z})$ through a flat
potential barrier of height $V$ grown along the $z$ [001] direction;
$\boldsymbol{k}_{\shortparallel}$ is the in-plane wave vector, and $k_{z}$ is
the wave vector along the growth direction, i.e., z axis. Then the electron
Hamiltonian including the spin-dependent $k^{3}$ Dresselhaus term is%

\begin{align}
H &  =\frac{P^{2}}{2m^{\ast}}+V(z)+H_{D}\text{,}\nonumber\\
H_{D} &  =\gamma_{i}[\sigma_{x}k_{x}(k_{y}^{2}-k_{z}^{2})+\sigma_{y}%
k_{y}(k_{z}^{2}-k_{x}^{2})+\sigma_{z}k_{z}(k_{x}^{2}-k_{y}^{2})]
\end{align}
where $m^{\ast}$ is the effective mass of electron, and $V(z)$ is the height
of barrier. $H_{D}$ describes the Dresselhaus spin-orbit coupling,
$\sigma_{\alpha}$ are the Pauli matrices, and $\gamma_{i}(i=1,2)$ describe the
strength of the Dresselhaus effect in the contact regions and the barrier. We
assume that the kinetic energy of the electron is substantially smaller than
the barrier height $V$ in the barrier \cite{INY,SAT} (we take $V=0.2eV$,
$E_{F}=0.02eV$ in our paper). The Hamiltonian in the barrier is simplified to%
\begin{equation}
H=\frac{P^{2}}{2m^{\ast}}+V(z)+\gamma_{2}(k_{x}\sigma_{x}-k_{y}\sigma
_{y})\frac{\partial^{2}}{\partial z^{2}}%
\end{equation}
The eigenvalues and eigenstates of the Hamiltonian are%
\begin{align}
E &  =\frac{\hslash^{2}(k_{\shortparallel}^{2}+q_{\pm}^{2})}{2m^{\ast}%
}+V(z)\pm\gamma_{2}k_{\shortparallel}q_{\pm}^{2}\text{,}\\
\chi_{\pm} &  =\frac{1}{\sqrt{2}}\binom{1}{\mp e^{-i\varphi}}\text{,}%
\end{align}
which correspond to the "+" and "-" electron states with opposite spin
orientation. The wave functions of the electron in the left source, the
barrier and right drain are%
\begin{align}
\Psi_{L} &  =\Psi_{i}+\exp(i\boldsymbol{k}_{\shortparallel}\cdot
\boldsymbol{\rho})%
%TCIMACRO{\dsum \limits_{j=\pm}}%
%BeginExpansion
{\displaystyle\sum\limits_{j=\pm}}
%EndExpansion
r_{j}\exp(-ik_{j}z)\chi_{j}\text{,}\nonumber\\
\Psi_{b} &  =\exp(i\boldsymbol{k}_{\shortparallel}\cdot\boldsymbol{\rho})%
%TCIMACRO{\dsum \limits_{j=\pm}}%
%BeginExpansion
{\displaystyle\sum\limits_{j=\pm}}
%EndExpansion
[A_{j}\exp(q_{j}z)+B_{j}\exp(-q_{j}z)]\chi_{j}\text{,}\nonumber\\
\Psi_{R} &  =\exp(i\boldsymbol{k}_{\shortparallel}\cdot\boldsymbol{\rho})%
%TCIMACRO{\dsum \limits_{j=\pm}}%
%BeginExpansion
{\displaystyle\sum\limits_{j=\pm}}
%EndExpansion
t_{j}\exp(ik_{j}z)\chi_{j}\text{,}%
\end{align}
here $\Psi_{i}=\exp(i\boldsymbol{k}_{\shortparallel}\cdot\boldsymbol{\rho})%
%TCIMACRO{\dsum \limits_{j=\pm}}%
%BeginExpansion
{\displaystyle\sum\limits_{j=\pm}}
%EndExpansion
\exp(ik_{j}z)\chi_{j}$ corresponds to the injected spin state of
electron\cite{INY,SAT}, $t_{\pm}$, $r_{\pm}$ are the transmission and
reflection coefficients for the spin states $\chi_{\pm}$, respectively.
$\boldsymbol{\rho}=(x,y)$ is the coordinate in the barrier plane, and
$\varphi$ is the polar angle of the wave vector $\boldsymbol{k}$ in the $xy$
plane. The wave vectors $q_{\pm}$($k_{\pm}$) inside and outside the barrier
are given by%
\begin{align}
q_{\pm} &  =\sqrt{\frac{\frac{2m_{2}^{\ast}V}{\hslash^{2}}+k_{\shortparallel
}^{2}-\frac{2m_{2}^{\ast}E_{F}}{\hslash^{2}}}{1\pm\frac{2m_{2}^{\ast}}%
{\hslash^{2}}\gamma_{2}k_{\shortparallel}}\text{,}}\nonumber\\
k_{\pm} &  =\sqrt{\frac{\frac{2m_{1}^{\ast}E_{F}}{\hslash^{2}}%
-k_{\shortparallel}^{2}}{1\pm\frac{2m_{1}^{\ast}}{\hslash^{2}}\gamma
_{1}k_{\shortparallel}A}}\text{,}%
\end{align}
where $A=\sqrt{1+\tan^{2}\theta+\tan^{2}\theta\sin^{2}2\varphi(\tan^{2}%
\theta/4-2)}$ and $m_{1}^{\ast}(m_{2}^{\ast})$ is the effective masses
outside(inside) the barrier. The boundary conditions including the Dresselhaus
effect are%
\begin{align}
\Psi_{L}|_{z=0} &  =\Psi_{b}|_{z=0}\text{,}\nonumber\\
\Psi_{b}|_{z=a} &  =\Psi_{R}|_{z=a}\text{,}\nonumber\\
j_{L}\Psi_{L}|_{z=0} &  =j_{b}\Psi_{b}|_{z=0}\text{,}\nonumber\\
j_{b}\Psi_{b}|_{z=a} &  =j_{R}\Psi_{R}|_{z=a}%
\end{align}
where $j_{i}(i=L,R,b)$ are the current operators in the left side, the right
side of the barrier, and inside the barrier,%
\begin{align}
j_{L,R} &  =\frac{1}{\hbar}\left(
\begin{array}
[c]{cc}%
-i\frac{\hslash^{2}}{m_{1}^{\ast}}\frac{\partial}{\partial z}+\gamma
_{1}k_{\shortparallel}^{2}\cos(2\varphi) & -2\gamma_{1}k_{\shortparallel
}e^{i\varphi}(-i\frac{\partial}{\partial z})\\
-2\gamma_{1}k_{\shortparallel}e^{-i\varphi}(-i\frac{\partial}{\partial z}) &
-i\frac{\hslash^{2}}{m_{1}^{\ast}}\frac{\partial}{\partial z}-\gamma
_{1}k_{\shortparallel}^{2}\cos(2\varphi)
\end{array}
\right)  \text{,}\nonumber\\
j_{b} &  =\frac{1}{\hbar}\left(
\begin{array}
[c]{cc}%
\frac{\hslash^{2}}{m_{2}^{\ast}} & -2\gamma_{2}k_{\shortparallel}e^{i\varphi
}\\
-2\gamma_{2}k_{\shortparallel}e^{-i\varphi} & \frac{\hslash^{2}}{m_{2}^{\ast}}%
\end{array}
\right)  (-i\frac{\partial}{\partial z})\text{,}%
\end{align}
Note that the off-diagonal elements of the current operators were neglected in
Ref.\cite{INY}. For the real case $m_{1}^{\ast}\gamma_{1}k_{\shortparallel
}/\hbar^{2}\ll1$ and $m_{1}^{\ast}\gamma_{1}k_{\shortparallel}^{2}%
\cos(2\varphi)/q_{\pm}/\hbar^{2}\ll1$, we obtain the analytical expression of
the spin-dependent transmission coefficient of the electron%
\begin{equation}
t_{\pm}=-\frac{4i\delta_{\pm}\delta_{\pm}^{\prime}k_{\pm}q_{\pm}e^{(-q_{\pm
}a-ik_{\pm}a)}}{(i\delta_{\pm}^{\prime}q_{\pm}-\delta_{\pm}k_{\pm}%
)^{2}e^{-2q_{\pm}a}-(i\delta_{\pm}^{\prime}q_{\pm}+\delta_{\pm}k_{\pm})^{2}%
}\text{,}%
\end{equation}
where $\delta_{\pm}=1\pm2m_{1}^{\ast}\gamma_{1}k_{\shortparallel}/\hbar^{2}$,
and $\delta_{\pm}^{\prime}=1\pm2m_{2}^{\ast}\gamma_{2}k_{\shortparallel}%
/\hbar^{2}$.

It is convenient to introduce the spin polarization $P$ determined by the
difference between the transmission of the spin states $\chi_{+}$ and
$\chi_{-}$,%
\begin{equation}
P=\frac{|t_{+}|^{2}-|t_{-}|^{2}}{|t_{+}|^{2}+|t_{-}|^{2}}\text{,}%
\end{equation}

The interface current due to spin-polarized electron transport through the
tunneling structure can be obtained through the spin density matrix\cite{SAT}%
\begin{equation}
\boldsymbol{j}_{\parallel}=e\sum_{\boldsymbol{k}_{\shortparallel},k_{z}>0}%
\tau_{p}Tr[\mathcal{T}\rho_{l}\mathcal{T}^{+}v_{z}]\boldsymbol{v}%
_{\shortparallel}\text{,}%
\end{equation}
where $\tau_{p}$ is the momentum relaxation time, $\rho_{l}$ is the electron
density matrix on the left side of the structure, and $\mathcal{T}$ is the
spin matrix of the tunneling transmission that links the incident spin wave
function $\Psi_{L}$ to the transmitted spin wave function $\Psi_{R}$,
$\Psi_{R}=\mathcal{T}\Psi_{L}$. The spin matrix of the electron transmission
through the structure is given by%
\begin{equation}
\mathcal{T}=\sum_{s=\pm}t_{s}\chi_{s}\chi_{s}^{+}\text{.}%
\end{equation}
In the case of small degree of spin polarization, the density matrix has the
form%
\begin{equation}
\rho_{l}=f_{0}I-\frac{df_{0}}{d\varepsilon}\frac{2p_{s}}{<1/\varepsilon
>}(\boldsymbol{n}_{s}\cdot\widehat{\sigma})\text{,}%
\end{equation}
where $f_{0}$ is the equilibrium distribution function of non-polarized
carriers, $p_{s}$ is the degree of the polarization, and $<1/\varepsilon>$ is
the average of the reciprocal kinetic energy of the carriers. $\boldsymbol{n}%
_{s}$ is the unit vector directed along the spin orientation, and the
orientations of spin $\boldsymbol{s}_{\pm}$ in the states "+" and "-" depend
on the in-plane wave vector of the electron and are given by $\boldsymbol{s}%
_{\pm}=(\mp\cos\varphi,\pm\sin\varphi,0)$. Taking into account the spin matrix
and the density matrix, the interface current is
\begin{equation}
j_{\parallel}=-C_{0}\sum_{\boldsymbol{k}_{\shortparallel},k_{z}>0}\frac
{df_{0}}{d\epsilon}[|t_{+}|^{2}-|t_{-}|^{2}]v_{z}v_{\shortparallel}\text{.}%
\end{equation}
where $C_{0}=e\tau_{p}\frac{p_{s}}{<1/\varepsilon>}$. It is interesting to
notice that the direction of the interface current depends on the spin
polarization of the injected electron $p_{s}$.

It is well know that normally the transmission through the barrier reaches
maximum for carriers propagating along the normal to the barrier in the
absence of spin-orbit interaction. But the spin-orbit coupling changes this
rule as shown in Fig. 1(see Eq. (9)). The tunneling transmission for the
spin-polarized electron with the finite in-plane wave vector $\boldsymbol{k}%
_{\parallel}$ is larger than that for the electron with the opposite in-plane
wave vector, $-\boldsymbol{k}_{\parallel}$. This asymmetry results in the
in-plane flow of the transmitted electron near the barrier, i.e., an interface
electric current.

For an electron with wave vector $\boldsymbol{k}$(see $\Psi_{L}$ in Eq. (5))
is injected into the barrier, the spin-dependent transmission is determined by
the width of the barrier and the Dresselhaus spin-orbit coupling strength(see
Eq. (8)). The spin-dependent transmission and the spin polarization $P$ are
plotted as a function of the barrier width $q_{0}a$ in Fig. 2. Here we
consider the Dresselhaus effect only in the barrier material. The material
parameter relevant to GaSb are used in our calculation $\gamma_{1}=0,$
$\gamma_{2}=187$ and the effective masses are $m^{\ast}=m_{2}^{\ast
}=0.041m_{0}$. It is shown that the spin-dependent transmission decreases
rapidly with increasing barrier width $q_{0}a$, while the spin polarization
increases gradually. From this figure, we can see that the results of
Ref.\cite{INY} agree with ours only in the thick barrier case, but there is a
big difference in the thin barrier case. From the numerical results, we can
find that the off-diagonal elements due to the Dresselhaus effect, in the
current operator play an important role in the spin-dependent transport,
especially in the thin barrier case.

Figure 3 describes the dependence of the magnitude of the interface current
$j_{\parallel}$ on the width of the barrier in unit of $C_{0}$ for incident
electrons, which form spin-polarized degenerate gas as in GaAs(Fig. 3(a)) and
in GaSb(Fig. 3(b)), respectively. From this figure, we can see that the two
results are very different. Our numerical results show that the maximum
interface current is overestimated in Ref.\cite{INY,SAT}, the two curves will
merge gradually together with increasing the barrier width. The numerical
accuracy spin-dependent tunneling and interface current could be improved
significantly by including the effect of the off-diagonal elements of the
current operator due to the Dresselhaus effect. The difference between the
present results and Ref.\cite{SAT} will increase with increasing $m_{2}^{\ast
}k_{\shortparallel}\gamma_{2}/\hslash^{2}$ comparing Fig. 3(a) and 3(b).

The spin polarization $P$ is plotted as a function of the barrier width $a$
for different strengths of the Dresselhaus effect in Fig. 4. From this figure,
we see that the spin polarization depends on the Dresselhaus effect
$\gamma_{i}$($i=1,2$) and the barrier width $a$. It is interesting to notice
that the spin polarization is enhanced by including the Dresselhaus effect
$\gamma_{1}$ in the contact regions, and increases with increasing the barrier
width $a$. On the other hand, the spin polarization decreases with diminishing
the strength of the Dresselhaus effect $\gamma_{2}$ in the barrier for a fixed
$\gamma_{1}$. The spin polarization saturates gradually with increasing
barrier width $a$ for $\gamma_{2}=0$, and reaches maximum when $\gamma
_{1}=\gamma_{2}=187$. These features can be understood from Fig. 4(b) and
4(c). The total transmission decreases with increasing barrier width and
almost the same for different strengths of the Dresselhaus effect(see Fig.
4(c)), but the differences between the spin states $\chi_{+}$ and $\chi_{-}$
reach the maxima for different strengths $\gamma_{1}$, $\gamma_{2}$(see Fig.
4(b)). Therefore the spin polarizations exhibit different behavior for
different strengths of Dresselhaus effect in the contact and barrier regions.
The spin polarization saturates gradually with increasing barrier width for
$\gamma_{1}=187$, $\gamma_{2}=0$ since the spin polarization is mainly caused
by the Dresselhaus effect $\gamma_{1}$ at both sides of the barrier, and
consequently almost independent of the barrier width. The spin polarization
approaches the maximum when $\gamma_{1}=\gamma_{2}$.

Fig. 5 shows the interface current $j_{\parallel}$ as a function of the
barrier width $a$ for different strengths of the Dresselhaus effect. From this
figure, we can see that the interface current $j_{\parallel}$ is enhanced by
the strength of the Dresselhaus effect $\gamma_{1}$ in the contact regions,
and show maxima at certain values of the barrier width $a$. It is interesting
to notice that the interface current $j_{\parallel}$ reaches maximum when
$\gamma_{1}=\gamma_{2}$, and decreases as the strength of the Dresselhaus
effect either in the barrier or in the contact regions decreases. Considering
the kinetic energy of the electron is substantially smaller than the barrier
height $V$ and $e^{-q_{\pm}a}\ll1$, we can obtain
\begin{equation}
j_{\parallel}\propto|t_{+}|^{2}-|t_{-}|^{2}\propto(\delta_{+}^{\prime}%
\delta_{+}e^{-2q_{+}a}-\delta_{-}^{\prime}\delta_{-}e^{-2q_{-}a})\text{.}%
\end{equation}
It is obvious that the interface current $j_{\parallel}$ increases with
increasing $\gamma_{2}$ if we fix $\gamma_{1}$. Similarly, the interface
current $j_{\parallel}$ decreases with decreasing $\gamma_{1}$ when we fix
$\gamma_{2}$. Therefore the interface current $j_{\parallel}$ reaches a
maximum when $\gamma_{1}=\gamma_{2}$.

In conclusion, we investigated the spin-dependent tunneling through a
symmetric semiconductor barrier consisting of zinc-blende semiconductor
material. It is interesting to notice that the spin polarization and the
interface current $j_{\parallel}$ are enhanced significantly by including the
Dresselhaus effect in the contact regions, and they all reach a maximum when
the strength of the Dresselhaus effect in the barrier is equal to that in the
contact regions, i.e., $\gamma_{1}=\gamma_{2}$.

\begin{acknowledgments}
This work was supported by City University of Hong Kong Strategic Research
Grant(No. 7001619) and the NSFC No. 60376016, 863 project No. 2002AA31, and
the special fund for Major State Basic Research Project No. G001CB3095 of China.
\end{acknowledgments}

\bigskip\newpage\begin{figure}[ptb]
\caption{The spin-dependent transmission of electron as a function of the
angle $\theta$ between the direction of $\boldsymbol{k}$ and the $z$ axis.}%
\ \end{figure}\begin{figure}[ptbptb]
\caption{The spin-dependent transmission of electron as a function of barrier
width $q_{0}a$ for GaSb material($\gamma_{1}=0$, $\gamma_{2}=187$, $V=0.2eV$).
The inset shows the spin polarization $P$ as a function of barrier width
$q_{0}a$ for the same condition.}%
\ \ \end{figure}\begin{figure}[ptbptbptb]
\caption{The interface current $j_{\parallel}$ as a function of barrier width
$a$ in unit of $C_{0}$ for (a) GaAs($\gamma_{1}=0,\gamma_{2}=24,V=0.2eV$) and
(b) GaSb($\gamma_{1}=0,\gamma_{2}=187,V=0.2eV$). The insets show the
difference in transmission between the spin states $\chi_{+}$ and $\chi_{-}$.}%
\ \ \end{figure}\ \ \begin{figure}[ptbptbptbptb]
\caption{(a) The spin polarization of electron as a function of barrier width
$a$ including the Dresselhaus effects inside and outside the barrier. (b) The
difference between the transmission of the spin states $\chi_{+}$ and
$\chi_{-}$. (c) The total of the transmission of the spin states $\chi_{+}$
and $\chi_{-}$.}%
\ \end{figure}\begin{figure}[ptbptbptbptbptb]
\caption{The interface current $j_{\parallel}$ as a function of barrier width
$a$ in unit of $C_{0}$ including the Dresselhaus effects inside and outside
the barrier.}%
\ \end{figure}

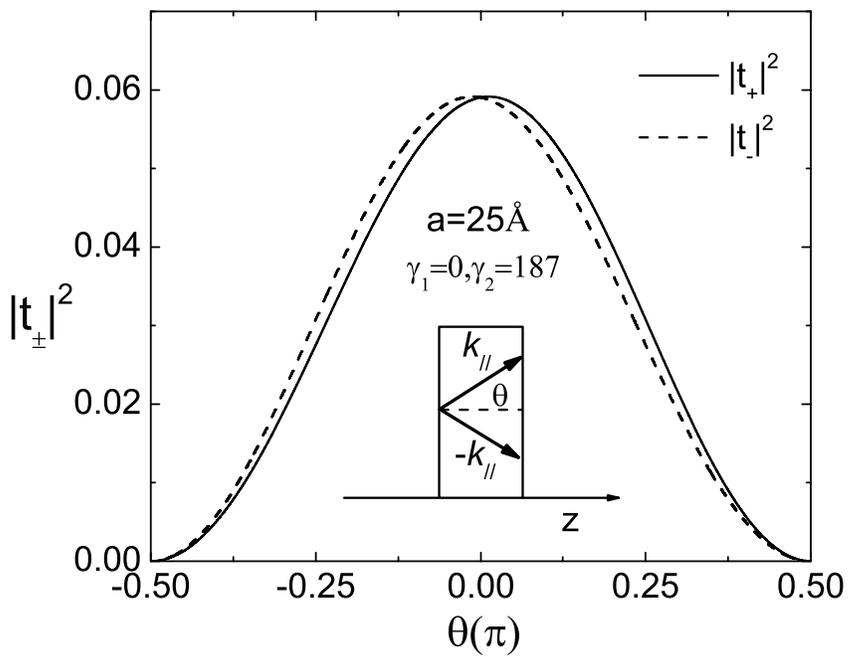
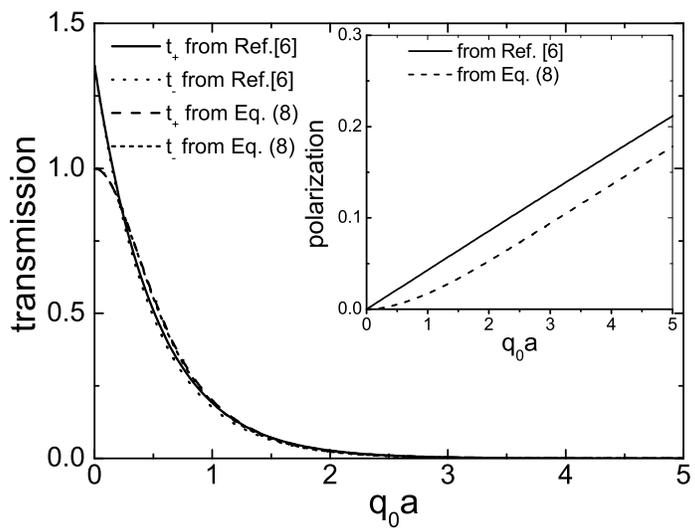
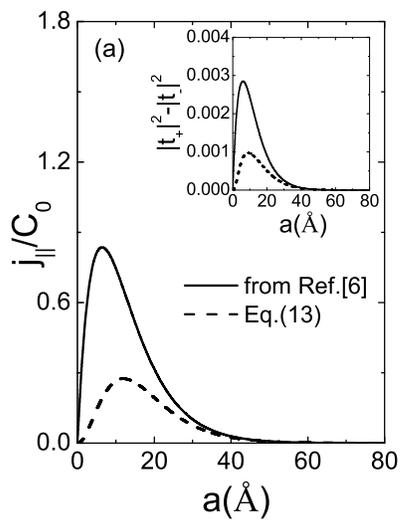
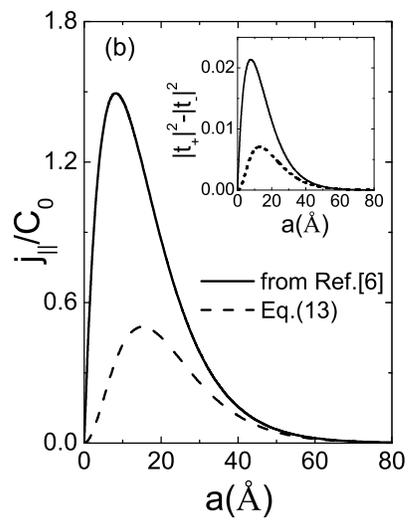
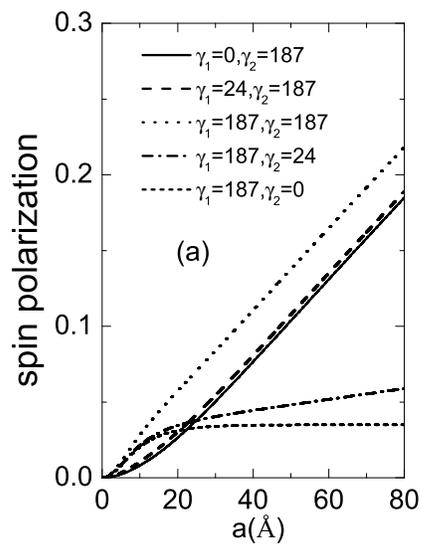
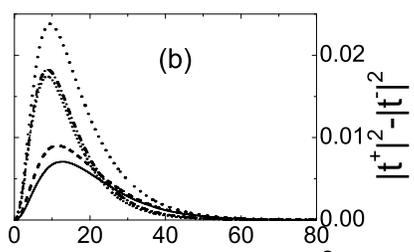
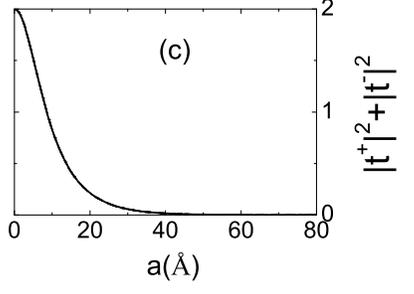
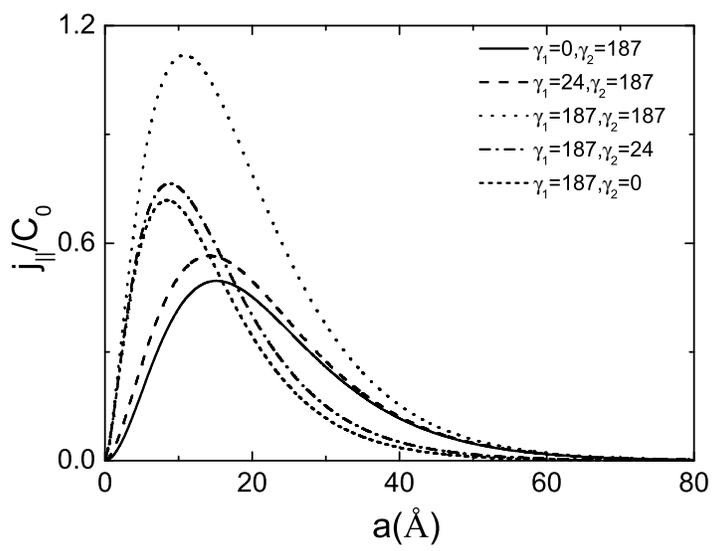
\end{document}